\documentstyle[12pt]{article}
\author{Salvatore Mignemi *, Hans-J\"urgen Schmidt **}
\title{Two-dimensional higher-derivative gravity and conformal
transformations
}
\date{}
\begin{document}
\maketitle

\bigskip

\centerline{
* Universit\'a di Cagliari, Dipartimento di Scienze Fisiche}
\centerline{I-09100 Cagliari, via Ospedale 72, Italy}

\bigskip

\centerline{
** Universit\"at Potsdam, Institut f\"ur Mathematik,
Projektgruppe  Kosmologie}
\centerline{
      D-14415 POTSDAM, PF 601553, Am Neuen Palais 10, Germany}

\bigskip

\begin{abstract}
We consider the lagrangian $L=F(R)$ in classical
(=non-quantized) two-dimensional fourth-order gravity and give
new relations to Einstein's theory with a non-minimally
coupled scalar field.

We distinguish between scale-invariant lagrangians and
scale-invariant field equations.
$L$ is scale-invariant for $F = c_1 R\sp {k+1}$ and a
divergence for $F=c_2 R$. The field equation is
scale-invariant not only for the sum of them, but also for
$F=R\ln R$. We prove this to be the only exception and show in
which sense it is the limit of
$\frac{1}{k} R\sp{k+1}$ as $k\rightarrow 0$. More generally:
Let $H$ be a divergence and $F$ a scale-invariant lagrangian,
then $L= H\ln F $ has a scale-invariant field equation.

Further, we comment on the known generalized Birkhoff theorem
and exact solutions including black holes.
\end{abstract}

PACS numbers: 04.20, 04.50

\section{Introduction}
 In recent years there has been a great interest in
two-dimensional theories of gravity [1-10], due in part to
their connection with string theories [11-15].
However, two-dimensional gravity models have a great interest
in themselves, since their qualitative features are similar to
those of general relativity, even if the mathematical
structure is much simpler. They can therefore be used to gain
some insight on the four-dimensional theory.

The essential property which distinguishes the 2-dimensional
theory from the higher-dimensional ones is the fact that the
Einstein-Hilbert lagrangian is a total derivative in two
dimensions. This problem is usually circumvented by
introducing a scalar field (sometimes called dilaton)
non-minimally coupled to the Ricci scalar [1, 2].

The action is however not uniquely defined in this way,
essentially because of the freedom in the choice of the
kinetic and potential terms for the scalar field. Thus one can
generate a large class of models, by simply requiring the
renormalizability of the theory [11, 12]. Some special
examples are given by the Jackiw-Teitelboim theory [1, 2], the
tree-level string lagrangian [13-15], and the 2-dimensional
limit of general relativity
 [16-18]. A one-parameter class of models with constant
potential  containing these special cases has been studied in
[8-9].

A different solution to the problem of defining a suitable
action for 2-dimensional gravity is given by higher derivative
theories. In this case one defines a lagrangian which is  a
non-linear function of the Ricci scalar, avoiding in this way
the problems found with the Einstein lagrangian in 2
dimensions [5, 6].

As is well-known, in dimensions higher than two,
higher-derivative models following from a non-linear
lagrangian $F(R)$ are conformally equivalent to general
relativity minimally coupled with a self-interacting scalar
field [19-23].

In two dimensions, since it is not possible to define a
minimally coupled theory, the situation is more subtle. The
existence of an equivalence between higher-derivative and
gravity-scalar theories has been noticed by several authors
[7, 10, 11, 28]. However, no general formulation of the
equivalence is available in the literature. Moreover, its
relation with conformal transformations of the metric has not
been stated explicitly.

In this paper, we give an explicit classification of the
gravity-scalar actions which are equivalent to
higher-derivative actions up to conformal transformations. The
existence of a non-trivial special case leads us to discuss
the nature of scale-invariance for two-dimensional theories.
Moreover, we briefly discuss the significance of the Birkhoff
theorem in this context and the black hole solutions of the
theory.

Some further discussion on different aspects of
two-dimensional gravity can be found [24-32]. In [33], also
two-dimensional gravity is considered, but they apply
independent variation with respect to metric and connection,
so the results are not directly comparable. In [34], there is
observed a universal behaviour in the process of forming a
two-dimensional black hole. Ref. [35] deals with the
evaporation of two-dimensional black holes, where $N$ scalar
fields have been added as source.

The paper is organized as follows:
in section 2 we review 2-dimensional higher-derivative
theories and discuss their connection with the more common
approach given by the addition of a non-minimally coupled
scalar field. Moreover, we study the action of a conformal
transformation on the lagrangian.
In section 3 we clarify the r\^ole of scale transformations
for the lagrangian and the field equations. Section 4 is
devoted to a review of the Birkhoff theorem in the context of
two-dimensional gravity. In section 5 we compare the exact
solutions of the theory in various gauges. We discuss the
results in the final section 6.

\bigskip

\section{Transformation from fourth to second order
}
\setcounter{equation}{0}

As is by now well known, higher-derivative gravity models in
dimensions $D>2$ can be reduced by means of a conformal
transformation to Einstein's theory minimally coupled to a
scalar field [19-23]. Consider for example the $D$-dimensional
action \begin{equation}
I = \int  L(R) \sqrt{\vert g \vert} d\sp D x
\end{equation}
where $L(R)=R\sp{k+1}$, $k \ne 0, \ -1 $ and $R\ne 0$. For
simplicity, we write the next formulas for the region $R>0$
only, the other sign gives analogous ones. If one defines the
scalar field $\sigma $ by
$$ e\sp{-2\sigma} \ \equiv \ \frac{dL}{dR} \ =  \
(k+1)R\sp k $$
and performs a conformal transformation
$$\tilde g_{ij} \ = \ e\sp{-2n\sigma} g_{ij}$$
where $n$ is a parameter to be fixed, one obtains the action
$$ I = \int e\sp{-[2+(D-2)n]\sigma} \ [
\tilde R + 2n(D-1) \tilde \nabla \sp 2 \sigma $$
$$ - n\sp 2 (D-1)(D-2) (\tilde \nabla \sigma )\sp 2 ] -
\Lambda \exp \{-(2\frac{k+1}{k} +
nD)\sigma  \}
\sqrt{\vert \tilde g \vert} d\sp D x $$
where $\Lambda = k/(k+1)\sp{1+1/k}$.
In particular, only if one chooses $n=-\frac{2}{D-2}$, the
scalar field is minimally coupled to the Einstein action as
follows:
$$ I = \int [
\tilde R  - \frac{D-1}{D-2} (\tilde \nabla \sigma )\sp 2 -
\Lambda \exp \{-2(\frac{k+1}{k} - \frac{D}{D-2})\sigma  \})
]
\sqrt{\vert \tilde g \vert} d\sp D x $$
This choice of $n$ is of course singular for dimension $D=2$.
This is due to the fact that in 2 dimensions $R\sqrt{g} $  is
a total derivative and therefore no analogue of the
 higher-dimensional minimally coupled action exists. It is in
fact necessary to make use of a non-minimally coupled scalar
field $\sigma $ and define an action of the kind
$$ \int e\sp{-2\sigma} \ R \ \sqrt{\vert g \vert }d\sp 2 x$$

Actually, in 2 dimensions, it is not even necessary to
perform a conformal transformation in order to get a linear
lagrangian from eq. (2.1). In fact, if one defines as before
$ e\sp{-2\sigma} \ \equiv \ \frac{dL}{dR} \ =  \
(k+1)R\sp k $ one gets
\begin{equation}
 I = \int [R
 e\sp{-2\sigma}  - \Lambda \exp \{-2\frac{k+1}{k}
\sigma  \}   ]
\sqrt{\vert  g \vert} d\sp 2 x
\end{equation}
This is a perfectly well defined action for 2-dimensional
gravity. If one performs a conformal transformation on eq.
(2.2) $\tilde g_{ij}  = e\sp{-2n\sigma} g_{ij}$, one gets
\begin{equation}
 I = \int e\sp{-2\sigma} \ [
\tilde R  + 4 n (\tilde \nabla \sigma )\sp 2 - \Lambda \exp
\{-2(\frac{1}{k}+n)\sigma  \}   ]
\sqrt{\vert \tilde g \vert} d\sp 2 x
\end{equation}
so that the gravitational part is unchanged, while the scalar
field acquires a kinetic term. All the actions (2.3) are
conformally equivalent in the sense that if $g_{ij}$ is a
stationary point of action (2.2) then  $\tilde g_{ij}$ is
one of (2.3). In particular, for $n=1$ one obtains the
well-known "string-like" action [7]:
$$
 I = \int e\sp{-2\sigma} \ [
\tilde R  + 4  (\tilde \nabla \sigma )\sp 2 - \Lambda
\exp \{-2(\frac{1}{k}+1)\sigma  \}
  ]
\sqrt{\vert \tilde g \vert} d\sp 2 x
$$
whose solutions are given by $\tilde g_{ij}  = e\sp{-2\sigma}
g_{ij}$.

The previous discussion can be generalized to the case when
the lagrangian is a generic function $L=F(R)$ of the
curvature. In this case, one defines $e\sp{-2\sigma} = G$
where $
G(R)=\frac{dF(R)}{dR}$ and the most general action related to
$L=F(R)$ by a conformal transformation
$\tilde g_{ij}  = e\sp{-2n\sigma} g_{ij}$
 of the two-dimensional metric takes the form
\begin{equation}
I=\int \{e\sp{-2\sigma} [\tilde R +
4n(\tilde \nabla \sigma )\sp 2 ] - V(\sigma) \}
\sqrt{\vert \tilde g \vert } d\sp 2 x
\end{equation}
where $n$ is a free parameter and
\begin{equation}
V(\sigma )=(RG - F) e \sp{-2n\sigma }
\end{equation}
For $n=0$, this is found in [28].

To conclude, we notice that
the action (2.2) admits two well-known theories as special
limiting cases. First, both for $k \longrightarrow \infty$ and

 for $k \longrightarrow \, - \, \infty$  it reduces to the
action of the Jackiw - Teitelboim theory
$$ I =\int \Phi [R - \Lambda] \sqrt{\vert  g \vert} d\sp 2 x
$$
where we have put $\Phi = e\sp{-2\sigma}$. Second, it can be
shown that the stationary points of (2.1) and (2.2) coincide
in the limit $k \longrightarrow 0$ with those of the
tree-level  string action. This limit is not at all trivial,
since for $k=0$, (2.1) is a total derivative, while (2.2) is
not defined.
As mentioned in [10, eq. (2.18)], the  $k \longrightarrow 0$
limit actually corresponds to the action
\begin{equation}
I = \int R \ln R \sqrt{\vert g \vert} d\sp 2 x
\end{equation}
This is not fully trivial but
can be understood starting from the well-known formula
\begin{equation}
\lim _{\epsilon \rightarrow 0} \frac{1}{\epsilon
} (e\sp{\epsilon x} - 1) \ = \ x
\end{equation}
We insert $x=\ln R$, multiply by $R$ and get
\begin{equation}
\lim _{\epsilon \rightarrow 0} \frac{1}{\epsilon
} (R\sp{\epsilon +1} - R) \ = \ R \ln R
\end{equation}
When inserted into the action (2.6), the $R$-term is a total
derivative, so one has
\begin{equation}
\int R \ln R \sqrt{\vert g \vert} d\sp 2 x
= boundary \ terms \ + \lim _{\epsilon \rightarrow 0}
\frac{1}{\epsilon
} \int R\sp{\epsilon +1}
\sqrt{\vert g \vert} d\sp 2 x
\end{equation}

There is an essential difference between the minimally and the
non-minimally coupled scalar field: If the kinetic term
$(\nabla \phi)\sp 2 $ is absent, then in the minimally coupled
case no dynamics for $\phi $ exists at all, whereas in the
non-minimally coupled case, the introduction of the kinetic
term does not alter the order of the corresponding field
equation. This is the reason for the possibility of actions
(2.2)/(2.3) becoming equivalent. In formulas: For $L=F(\Phi ,
R)$ with $G=\frac{\partial F}{\partial R}$ one gets
$0=\frac{\partial F}{\partial \Phi}$, $F=GR + \Box G$ and the
trace-free  part of
$G_{;ij}$ has to vanish. In the non-minimally coupled case,
 i.e. $\frac{\partial G}{\partial \Phi} \ne 0$, the parts with
$G_{;ij}$ contain the dynamics for $\Phi $.

\bigskip

\section{On different notions of scale-invariance
}
\setcounter{equation}{0}

The lagrangian density $R \ln R$ eq. (2.8) possesses also some
peculiar properties in relation with the scale invariance of
the theory.
For the notion of scale-invariance one has to specify to which
situation it refers. Here, we distinguish two different
notions for the following situation: We consider a
 two-dimensional Riemannian or Pseudoriemannian metric
$g_{ij}$ with curvature scalar $R$. Let a Lagrangian $L=F(R)$
be given where $F$ is a sufficiently smooth function (three
times differentiable is enough) and $G=\frac{dF}{dR}$. The
variational derivative of $L\sqrt{\vert g \vert }$ with
respect to
 $g_{ij}$ gives a fourth order field equation. The trace of
that equation reads
\begin{equation}
0 \ = \ G R - F + \Box G
\end{equation}
The field equation is completed by requiring that the
trace-free part of $G_{;ij}$ vanishes.

First definition: Let $\alpha $ be an arbitrary constant and
 let $\tilde g_{ij} \ = \ e\sp{2\alpha} g_{ij}$. Then, e.g.,
$\tilde R = e\sp{-2\alpha}R$ etc. The Lagrangian $L$ is called
scale-invariant if there exists a function $f(\alpha )$ such
that for all metrics it holds $$\tilde L =
f(\alpha ) \, L$$

One can get some knowledge on the function $f$ as follows:
 We apply the defining condition with $\beta$ instead of
$\alpha $ and with
$\hat g_{ij} \ = \ e\sp{2\beta} \tilde g_{ij}$. This leads to
\begin{equation}
\hat L = f(\beta) \tilde L = f(\beta)f(\alpha)L =
f(\beta + \alpha)L
\end{equation}
This last equality can be fulfilled only if there exists a
constant real number $m$ such that
$$f(\alpha) = e\sp{2\alpha m}$$
Our definition is therefore equivalent to:

The Lagrangian $L$ is called scale-invariant if there exists a
constant $m$ such that for all metrics it holds
\begin{equation}
\tilde L =
e\sp{2\alpha m} \, L
\end{equation}

We want to find out all scale-invariant Lagrangians. To this
end we insert $R=1$ into eq. (3.3), i.e. into
$$F(e\sp{-2\alpha}R) = e\sp{2\alpha m}F(R)$$
We use $x=e\sp{-2\alpha}$ and $c=F(1)$.
We get $F(x)=c x\sp{-m}$. This is the sense in which usually
$L=R\sp{k+1}$
is called the scale-invariant gravitational Lagrangian.

$L$ is a divergence iff the field equation is identically
fulfilled. For the situation considered here this takes place
if and only if $F(R) = cR$ with a constant
 $c$.

Let us now come to the second definition: Let $\alpha $ be an
arbitrary constant and
 let $\tilde g_{ij} \ = \ e\sp{2\alpha} g_{ij}$.
 The field equation following from the Lagrangian $L$ is
called scale-invariant if there exist  functions $f(\alpha )$
and   $g(\alpha )$ such that for all metrics it holds $$\tilde
L =
f(\alpha ) \, L \ + \ g(\alpha ) R$$

This definition is equivalent to: The field equation following
from $L$ is called scale-invariant iff $L$ is scale-invariant
up to a divergence. It is motivated by the fact that for a
 scale-invariant field equation and one of its solutions
$g_{ij}$, the homothetically transformed  $\tilde g_{ij}$ is a
solution, too.

To find out all scale-invariant field equations, we write the
analogue to eq. (3.2), i.e.
$$0=[f(\alpha + \beta) \ - \ f(\alpha )f( \beta) ] F(R) \ + $$
$$[g(\alpha + \beta) \ - \ g(\alpha )f( \beta)
\ - \ g(\beta)e\sp{-2\alpha}
 ] R  $$
A linear function $F(R)$ gives always rise to a
 scale-invariant field equation. For non-linear functions
$F(R)$, however, both lines of the above equation must vanish
separately. The vanishing   of the first line gives again
$f(\alpha) = e\sp{2\alpha m}$. We insert this into the second
line and get
$$g(\alpha + \beta) = g(\alpha ) e\sp{2m\beta} +
g(\beta)e\sp{-2\alpha}
  $$
To solve this equation it proves useful to define
$h(\alpha)=g(\alpha)e\sp{2\alpha}$ leading to
$$h(\alpha + \beta) = h(\alpha ) e\sp{2(m+1)\beta} + h(\beta)
  $$

1. case: $m\ne -1$: After some calculus one gets
$F(R) = cR\sp{-m} + kR$, just the expected sum.

2. case: $m = -1$: Then there exists a constant $c$ such that
$h(\alpha)=c \cdot \alpha$, i.e.,
$g(\alpha)=c \cdot \alpha e\sp{-2\alpha}$. To find the
corresponding $F(R)$ we have to solve
$$F(e\sp{-2\alpha}R)=e\sp{-2\alpha}[F(R)+c\alpha R]
$$
which is done by
\begin{equation}
F(R) \ = \ - \frac{c}{2} R \ln R + kR
\end{equation}
$k$ being a constant. So we see: $L= R \ln R$ is not a
 scale-invariant lagrangian but it has a scale-invariant
 field equation and one learns: To find out all lagrangians
being "scale-invariant up to a divergence" it does not suffice

to add all possible divergencies (here: $kR$, $k$ being
constant) to all scale-invariant lagrangians (here:
$L = c R\sp{-m}$).

The distinction made here can analogously be formulated for
higher dimensions. One gets the following: Let $H$ be a
divergence and $F$ be a scale-invariant lagrangian, then
$L=H\ln F$ gives rise to a scale-invariant field equation.
This covers the above example for $D=2$ with $H=F=R$.

One might have got the impression that if a scale-invariant
lagrangian is rewritten with a conformally transformed metric
then the resulting field equation remains essentially the
same. But this is not always the case. The typical example is:
Take the Einstein - Hilbert action $I=\int R\sqrt{\vert g
\vert } d \sp D x$ and define
 $\hat g_{ij} = R\sp m g_{ij}$ for $R>0$. Then
$ \sqrt{\vert \hat  g \vert } = R\sp{Dm/2} \sqrt{\vert g \vert
} $ . For $Dm=2$, $I=\int \sqrt{\vert \hat g \vert }d\sp D x$,
so only for $Dm \ne 2$ the corresponding field equations
become equivalent.

\bigskip

\section{The generalized Birkhoff theorem
}
\setcounter{equation}{0}

In  [5] and [11] the following was shown: Let $L=F(R)$ be a
non-linear  Lagrangian in two dimensions and $G=\frac{dF}{dR}
$; then
$\Theta \sp i = \epsilon \sp {ij} G_{ ;j} $ is a Killing
vector. This result is called "generalized Birkhoff theorem"
for its type being "a  spherically symmetric vacuum solution
has an additional  Killing  vector"; in fact, in one spatial
dimension, the assumption of  spherical symmetry is empty.

To know whether the existence of a Killing vector implies a
local symmetry, one must be sure that it does not vanish.
Supposed, $\Theta \sp i$ identically vanishes, then $G$ must
be a  constant, and so $R$ is a constant. Then the space is of
constant  curvature and a non-vanishing non-lightlike Killing
vector exists. Supposed,
 $\Theta \sp i$ is a non-vanishing null vector, then again,
the  space turns out to be of constant curvature. So the only
possibility for  $\Theta \sp i$ becoming lightlike is at a
line  (the horizon) where it changes its signature. These are
the  solutions being known under the name "two-dimensional
black  holes".

Generically (meaning here: in a region where the  Killing
vector
is  non-lightlike) one can always write the solution as
\begin{equation}
ds \sp 2 = A\sp 2(x)dx\sp 2 \pm B\sp 2(x)dy\sp 2
\end{equation}
As usual, the free transformation of $x$ can be used to
eliminate  $A$ or $B$; especially the condition $AB=1$ leads
to generalized  Schwarzschild coordinates.
What is essential for eq. (4.1): The change between Euclidean
and  Lorentzian signature is possible by the complex rotation
$y \longrightarrow iy $. This is of course only local and
generically, so that the global topology may be (and indeed,
is)  different, but in higher dimensions such a relation does
not need  to take place even locally. (The reason is: in two
dimensions, a  Killing vector is automatically
hypersurface-orthogonal.)

This generalized Birkhoff theorem has the consequence that
special solutions having symmetries (see sct. 5 below) found
in  the past already cover the whole space of solutions. Of
course, the theorem can be
extended to the gravity-scalar theories, owing to their
equivalence with higher-derivative theories.

\bigskip

\section{Exact solutions
}
\setcounter{equation}{0}
The solution of the field equations stemming from eqs.
(2.1)/(2.3) have been found in [5, 6] and, in a conformal
gauge, in [7].
We shortly discuss them in this section.

For $k\ne -1/2$, the Lorentzian signature solutions can be
written in the so-called Schwarzschild gauge as [5, 6]:
\begin{equation}
ds\sp 2=-A\sp 2(x)dt\sp 2+A\sp {-2}(x)dx\sp 2
\end{equation}
with
$$A\sp 2(x)=-C+ \vert x \vert \sp {2+1/k}$$
while for $k=-1/2$,
$$A\sp 2(x)=-C+\ln\vert x\vert $$
where $C$ is a free parameter, proportional to the mass of the
solution. In particular, for positive $C$ one gets in general
black hole solutions, while for negative $C$ one has naked
singularities. $C=0$ corresponds to the self-similar ground
state of the theory.
The conformal gauge solutions found in [7] can be obtained
from (5.1) for $k \ne - 1/2$ by the coordinate transformation
$$\rho=\int dx [ -C+ \vert x \vert \sp {2+1/k}] \sp{-1}$$
In particular, if $C=0$, $\rho=x\sp {-(1+1/k)}$, and
$$ds\sp 2= \rho\sp {-(2k+1)/(k+1)}(d\rho\sp 2-dt\sp 2).$$

Let us discuss in some detail the properties of the solutions:
in the Schwarzschild gauge the curvature is simply given by:
$R=- d\sp 2 (A\sp 2)/dx\sp 2$. Thus one sees that a
singularity
 (in the sense of a diverging curvature scalar $R$)
is present at the origin only if $k$ is negative. Moreover,
for positive $C$, a horizon is present at $x=C\sp {k/(k+2)}$
for any $k$. The horizon is absent if $C$ is negative.

The asymptotic properties of the solutions  are also
 interesting: for negative $k$ the curvature vanishes at
infinity, but only in the limit case $k=0$ the solutions are
asymptotically flat in the usual sense (i.e. $A\rightarrow 1$
at infinity). For positive $k$ the curvature diverges at
infinity. Finally, in the limit $k\rightarrow\pm\infty$
(Jackiw-Teitelboim theory), the solutions are asymptotically
anti-de-Sitter.

The limit case $k\rightarrow 0$ has been studied in [5]. In
this case the solutions coincide with the "stringy" solutions
found in [13, 14] and with a solution of Liouville gravity
(see [30-32] for details):
$$A\sp 2(x)=1-Ce\sp x$$
and describe asymptotically flat black holes.

To summarize, regular black hole solutions are found only for
$k\le 0$ and positive $C$.

In a similar manner, one can discuss the solutions of the
Euclidean theory. Apart from the horizon, theses are simply
obtained by setting $t \longrightarrow it$. In the black hole
case, the conical singularity at the origin (i.e., the point
corresponding to the horizon) can be removed
 by a standard procedure, requiring that the Euclidean time
has periodicity $\beta$ which is related to the temperature
$T$ of the black hole via
$$T=\beta \sp {-1} = \frac{2k+1}{4\pi k} C\sp{(k+1)/(2k+1)}$$

\bigskip

\section{Discussion
}
In this paper, we considered several types of two-dimensional
theories of gravity. We restricted to the classical (=
non-quantum) case; the metric and one scalar field are the
only ingredients (no torsion, no further matter). The aim of
the paper was to clarify the conformal relation between
different versions of the theory; especially, we carefully
distinguished between transformations on the lagrangian and on
the field equation's level.

Section 2 dealt with the conformal transformation from a
non-linear lagrangian $L(R)$ (corresponding to a fourth-order
field equation) to Einstein's theory with one additional
scalar field. To simplify the formulas we first considered the
case $L(R)=R\sp{k+1}$, ($k\ne 0, -1$) to show how the
transformation breaks down for dimension $D=2$ if the scalar
field is required to be minimally coupled. The reason is that
for $D=2$ the curvature scalar is a divergence. So, for $D=2$,
the conformal equivalence becomes possible for a non-minimally
coupled scalar field only. We showed this in two steps: first
for $L(R)=R\sp{k+1}$, and second, eqs. (2.4, 2.5), the
one-parameter set (the parameter is $n$) of conformal
transformations from a general non-linear $L(R)$ to Einstein's
theory with a non-minimally coupled scalar field. (The points
where this transformation becomes singular are not explicitly
written down but become clear from the formulas.)
Only few special cases of this result can be found in the
literature. From eq. (2.4) it becomes clear that the kinetic
term of the scalar field vanishes for $n=0$. This does not
destroy the equivalence because the dynamics of the scalar
field now comes from the non-minimal coupling to $R$. So, the
change from $n=0$ to $n\ne 0$ represents a conformal
transformation of a scalar field without to a scalar field
with kinetic term. This generalizes the class of conformal
transformation of [12] relating between
$$L=\frac{1}{2} (\nabla \Phi )\sp 2 + F(\Phi )R+U(\Phi)$$
and
$$L=\frac{1}{2} (\nabla \phi )\sp 2 + \frac{q \phi }{2}
R+V(\phi)$$

	To avoid possible misunderstandings: Some papers do not
have the factor 4 in front of the kinetic term as we have. In
[10], e.g., one has
$$L = e\sp{\Phi}[ R + (\nabla \Phi )\sp 2 + \lambda ] $$
If one inserts $\Phi = \pm 2 \phi $ then one gets
$$L = e\sp{\pm 2 \phi}[ R + 4(\nabla \phi )\sp 2 + \lambda ]
$$
so this is only a notational difference. A further
misunderstanding can appear by noting that $R\sp{k+1}$ tends
to $R \ln R$ as $k \longrightarrow 0$. In eqs. (2.6 - 2.9) we
clarified in which sense this is a mathematically correct
statement.

In section 3 we distinguished different notions of
scale-invariance. It turned out that two of them are
essentially different: Scale-invariant lagrangians and
scale-invariant field equations. It is trivial to see that the
sum of a scale-invariant lagrangian and an arbitrary
divergence gives rise to a scale-invariant field equation.
Surprisingly, these sums do not yield all scale-invariant
field equations. One (the only !) counterexample is the often
discussed case $L=R \ln R$.

In section 4 we discussed the fact that in the models under
consideration a non-vanishing Killing vector always exists
(generalized Birkhoff theorem). Here we want to emphasize:
A) that this does not need the scale-invariance of the action
(a case for which it is often formulated) but that it takes
place for all models. B) The conformal transformation shows
that the Birkhoff theorem is valid in all the versions of
two-dimensional gravity under consideration, and C) it is just
this Birkhoff theorem which makes possible (at least locally)
the complex rotation from Euclidean signature to Lorentz
signature solutions; the latter are discussed as
two-dimensional black holes. Section 5 represents known exact
solutions in a better readable form.

\bigskip

{\it Acknowledgement}. We thank C Gundlach, S Kluske, M Rainer
and S Reuter  for valuable comments. H-J S acknowledges
financial support from the
 Wissenschaftler-Integrations-Programm under contract Nr.
015373/E and from the Deutsche Forschungsgemeinschaft under
Nr. Schm 911/5-1. SM wishes to thank MRST for financial
support and the Institute of Mathematics of Potsdam University
for kind hospitality.

\bigskip

{\it References}

\bigskip

[1] Jackiw R 1984  in {\sl Quantum Theory of gravity},
Christensen S M ed. (Adam Hilger, Bristol) p. 403

[2] Teitelboim C 1983 Phys. Lett. {\bf B 126} 41

[3] Mann R B, Shiekh A and Tarasov I 1990 Nucl. Phys. {\bf B
341} 134

[4] Mann R B 1992 Gen. Rel. Grav. {\bf 24} 433

[5] Schmidt H-J 1991 J. Math. Phys. {\bf 32} 1562

[6] Schmidt H-J 1992, p. 330 in:
Relativistic Astrophysics and Cosmology, Eds.: Gottl\"ober S
M\"ucket J and M\"uller V, World Scientific, Singapore

[7] Mignemi S 1994 Phys. Rev. {\bf D 50} R4733

[8] Lemos J P L and S\'a P M 1994 Phys. Rev. {\bf D 49} 2997

[9] Cadoni M and Mignemi S 1994 preprints INFN-CA-20-93 and
INFNCA-TH-94-4

[10] Frolov V P 1992 Phys. Rev. {\bf D 46} 5383

[11] Banks T and Loughlin M O 1991 Nucl. Phys. {\bf B 362} 649

[12] Russo J G and Tseytlin A A 1992 Nucl. Phys. {\bf B 382}
259

[13] Witten E 1991 Phys. Rev. {\bf D 44} 314

[14] Mandal G, Sengupta A M and Wadia S R 1991 Mod. Phys.
Lett. {\bf A 6} 1685

[15] Callan C G, Giddings S B, Harvey J and Strominger A 1992
Phys. Rev. {\bf D 45} 1005

[16] Sikkema A E and Mann R B 1991 Class. Quantum Grav. {\bf
8} 219

[17] Mann R B  and  Ross S F 1993 Class. Quantum Grav. {\bf
10} 1405

[18] Lemos J P L and S\'a P M 1994 Class. Quantum Grav. {\bf
11}
 L11

[19] H.-J. Schmidt 1987 Astron. Nachr. {\bf 308} 183

[20] Magnano G, Ferraris M and Francaviglia M 1987 Gen. Rel.
Grav. {\bf 19} 465

[21] Barrow J and Cotsakis S 1988 Phys. Lett. {\bf B 214} 515

[22] Schmidt H-J 1988 Astron. Nachr. {\bf 309} 307

[23] Maeda K 1989 Phys. Rev. {\bf D 39} 3159

[24] Katanaev M O, Volovich I V 1990 Ann. Phys. (NY) {\bf 197}
1

[25] Kummer W and Schwarz D 1992 Nucl. Phys. B {\bf 382} 171

[26] Grosse H, Kummer W, Presnajder P and Schwarz D 1992
J. Math. Phys. {\bf 33} 3892

[27] Nojiri S and Oda I 1994 Mod. Phys. Lett. {\bf A 9} 959

[28] Solodukhin S N  1994 preprint JINR E2-94-185

[29] Gegenberg J, Kunstatter G and Louis-Martinez D 1994
preprint gr-qc/9408015.

[30] Christensen D and Mann R B 1992 Class. Quantum Grav. {\bf
9} 1769

[31] Mann R B  and  Ross S F 1992 Class. Quantum Grav. {\bf 9}
2335

[32] Mann R B, Morris M  and  Ross S F 1993 Class. Quantum
Grav. {\bf 10} 1477

[33] Ferraris M, Francaviglia M and Volovich I 1994 Class.
Quantum Grav. {\bf 11} 1505

[34] Strominger A and Thorlacius L 1994 Phys. Rev. Lett. {\bf
72} 1584

[35] Hawking S 1992 Phys. Rev. Lett. {\bf 69} 406
\end{document}